\documentclass[preprint,preprintnumbers,amsmath,amssymb]{revtex4}
 \usepackage{dcolumn}
 \usepackage{bm}
 \usepackage{graphicx}
 \usepackage{subfigure}
 \usepackage{color}
 \begin{document}

 \title{Hawking radiation, local temperatures, and nonequilibrium thermodynamics of the black holes with non-killing horizon}

 \author{Ran Li$^{1,2}$}
 \thanks{liran@htu.edu.cn}

 \author{Jin Wang$^{2,3}$}
 \thanks{Corresponding author: jin.wang.1@stonybrook.edu}

 \affiliation{
 $^1$ School of Physics, Henan Normal University, Xinxiang 453007, China\\
 $^2$ Department of Chemistry, SUNY at Stony Brook, Stony Brook, New York 11794, USA\\
 $^3$ Department of Physics and Astronomy, SUNY at Stony Brook, Stony Brook, New York 11794, USA\\}

 \begin{abstract}

Recently, a class of stationary black hole solutions with non-killing horizon in the asymptotic AdS bulk space (i.e. non-equilibrium black funnel) was constructed to describe the far from equilibrium heat transport and particle transport from the boundary black holes via AdS/CFT correspondence. It is generally believed that the temperature of a black hole with non-killing horizon can not be properly defined by the conventional methods used in the equilibrium black holes with killing horizon. In this study, we calculate the spectrum of Hawking radiation of the non-equilibrium black funnel using the Damour-Ruffini method. Our results indicate that the spectrum and the temperatures as well as the chemical potentials of the non-equilibrium black funnel do depend on one of the spatial coordinates. This is different from the equilibrium black holes with killing horizon, where the temperatures are uniform. Therefore, the black hole with non-killing horizon can be overall in non-equilibrium steady state while the Hawking temperature of the black funnel can be viewed as the local temperature and the corresponding Hawking radiation can be regarded as being in the local equilibrium with the horizon of the black funnel. By AdS/CFT, we discuss some possible implications of our results of local Hawking temperature for the non-equilibrium thermodynamics of dual conformal field theory. We further discuss the nonequilibrium thermodynamics of the black funnel, where the first law can be formulated as the entropy production rate being equal to the sum of the changes of the entropies from the system (black funnel) and environments while the second law is given by the entropy production being larger than or equal to zero. We found the time arrow emerged from the nonequilibrium black hole heat and particle transport dissipation. We also discuss how the nonequilibrium dissipation may influence the evaporation process of the black funnel.
 \end{abstract}



 \maketitle

 \section{Introduction}

The uniqueness theorem \cite{HawkingCMP,HawkingGR,Wald} states that all regular stationary, asymptotically flat non-degenerated black hole solutions of the Einstein-Maxwell equations of gravitation and electromagnetism in four dimensional general relativity can be uniquely characterized by their mass, electric charge, and angular momentum and have compact horizon
topology of $S^2$. The most general solution to the vacuum Einstein-Maxwell equations is characterized by the 3-parameter Kerr-Newman family. It seems that all the information about the matter that formed a black hole is inaccessible to the external observer. Wheeler used the phrase "black holes have no hair" to express the idea of the uniqueness theorem \cite{Misner}.

In recent years, in the context of AdS/CFT correspondence \cite{Maldacena,GKP,Witten}, many hairy black holes that evade the assumptions of uniqueness theorem have been constructed and used to describe the strongly coupled properties of condensed matter systems \cite{HLS,DSW}. For example, in holographic superconductor models \cite{HHH}, scalar, vector, and tensor fields can be added to the Einstein-Hilbert action with negative cosmological constant to construct the bulk spacetime dual to the boundary superconducting system. These examples indicate that the uniqueness theorem can be evaded in AdS asymptotics.

Another way to evade the assumptions of uniqueness theorem is to construct the black hole solutions with the bulk horizons that extend to the asymptotic regions where the boundary conditions can be imposed. This programm first initiated in \cite{HMR} for the aim of gaining some insight into the strong coupling properties of quantum field in curved spacetime, and continued in the following studies \cite{HMRCGQ,CDMS,FLW,FM,SW,FMS,FW,FS,SWJhep}. (See \cite{MRW} for a review.) According to the the connectedness of the bulk horizon(s), the solutions can roughly be divided into two classes: black funnels and black droplets.

Black funnel solution has a single connected bulk horizon that extends to meet the conformal boundary.
Then the induced conformal boundary has smooth horizons as well. In particular, in \cite{FMS}, the black funnel solution in vacuum Einstein-Hilbert gravity with negative cosmological constant was numerically constructed, where the boundary spacetime contains a pair of black holes connected through the bulk by a tubular bulk horizon. The boundary conditions that the boundary black holes have two different temperatures can be imposed. These are examples of the stationary bulk black hole with a non-Killing horizon, which can be used to describe the far from equilibrium transport of heat on the boundary from the hotter boundary black hole to the cooler boundary black hole by using the holographic dictionary.

The well known black hole thermodynamics told us that the surface gravity $\kappa$ is uniform everywhere on the killing horizon. This indicates that killing horizon can be viewed as being in thermal equilibrium and the temperature is well-defined by Hawking's result $T=\frac{\kappa}{2\pi}$.
The surface gravity and the temperature of the black hole with the non-killing horizon can not be properly defined by the conventional methods used in the black hole with killing horizon.
For example, Visser has calculated the surface gravity of the black holes with non-killing horizon, which shows the different definitions of surface gravity can lead to different results \cite{Visser}.
From the thermodynamics perspective, black holes with killing horizon can be described by the equilibrium thermodynamics. The temperature of the equilibrium black hole is related to the surface gravity of the horizon while the entropy is found to be proportional to the area of the horizon.
The thermodynamics of equilibrium black hole can be formulated as the internal energy change of the black hole being equal to the heat generated (product of temperature and entropy change). For non-killing horizon, the temperature of the black hole can not be treated as constant. The equilibrium description of the black hole is no longer valid in this situation. An inhomogenous temperature is expected to lead to heat flow and generate thermodynamic dissipation. Therefore, the thermodynamics of this kind of non-killing black holes is important to describe the overall behavior and should have nonequilibrium nature, but has not been formulated yet. Furthermore, the nonequilibrium evaporation dynamics of the non-killing black hole is also not discussed so far.

In this study, we will address how to calculate the Hawking temperature of non-killing horizon of stationary black hole solution. We take the previous mentioned black funnel solution as an example.
For the black funnel solution, the bulk horizon is non-compact, and can be extended to conformal boundary. From the thermodynamic point of view, the heat, which is originated from the boundary black holes with different temperatures, can be transported on the bulk horizon. Then the horizon of the black funnel no longer has a constant temperature. This means that the non-killing horizon should be taken as a non-equilibrium thermodynamic system.

There are several methods to derive Hawking radiation in literatures, for example, quantum field in curved space used by Hawking \cite{Hawking}, Damour-Ruffini method \cite{DR}, anomaly method \cite{CF}, and quantum tunneling method \cite{PW}, et al. The Damour-Ruffini method \cite{DR} has been generalized by Zhao et al to study the Hawking radiation from the dynamical black holes and accelerating black holes \cite{ZD}. We find that the Damour-Ruffini method is a convenient method to calculate the radiation spectrum of the non-killing horizon. By properly investigating the behavior of a scalar field near the horizon, we are able to derive the local temperature and the corresponding radiation spectrum of the bulk horizon of the black funnel. Our results indicate that the spectrum and the temperatures as well as the chemical potentials of non-equilibrium black funnel do depend on one of the spatial coordinates, which is different from the conventional case of the constant temperature equilibrium black hole with killing horizon. Therefore, the black hole with non-killing horizon can be overall in non-equilibrium steady state while the Hawking temperature of black funnel can be viewed as the local temperature and the corresponding Hawking radiation can be regarded as being in local equilibrium with the horizon of the black funnel. By AdS/CFT, we discuss some possible implications of our results of the local Hawking temperature for the non-equilibrium thermodynamics of dual conformal field theory. We further discuss the nonequilibrium thermodynamics of the steady state black funnel. We suggest the first and second laws of the thermodynamics of the noequilibrium steady state black funnel. We discuss the time arrow emerged from the nonequilibrium black hole heat and particle transport dissipation. We also consider the effect of the nonequilibrium thermal transport in addition to Hawking radiation on the evaporation process of black funnel.

 \section{Hawking Radiation of the Nonequilibrium Black Hole-Black Funnel}

 According to AdS/CFT correspondence \cite{Maldacena,GKP,Witten}, the conformal filed theory on the boundary is equivalent to the AdS spacetime in the bulk. A conformal field theory with thermal temperature is shown to be equivalent to an AdS black hole (black brane) in the bulk. All these are based on the equilibrium description. One can naturally ask the equation what the corresponding bulk spacetme is if the boundary conformal fluid is under nonequilibrium condition (for example, driven by a temperature gradient). The earlier exploration focused on initial temperature contrast setup and the subsequent evolution, where the bulk spacetime becomes a boosted black brane \cite{BDLS}. More recent work leads to the more fixed nonequilibrium setup by imposing two black holes on the boundary edges. As a result, the bulk spacetime has two outcomes: steady state black hole with non-killing horizon named black funnel and equilibrium black hole with killing horizon named black droplet. It is the purpose of this work to quantify the Hawking radiation and local temperature of the black hole with non-killing horizon. We will also formulate the non-equilibrium thermodynamics and the evaporation dynamics of the non-killing black holes.

\subsection{Nonequilibrium Black Hole: Black funnel}

In this section, we want to calculate the spectrum of Hawking radiation from the bulk horizon of the black funnel solution based on the Damour-Ruffini method \cite{DR}. For this aim, let us briefly introduce the construction of black funnel solution.

\begin{figure}
  \centering
  \includegraphics[width=7cm]{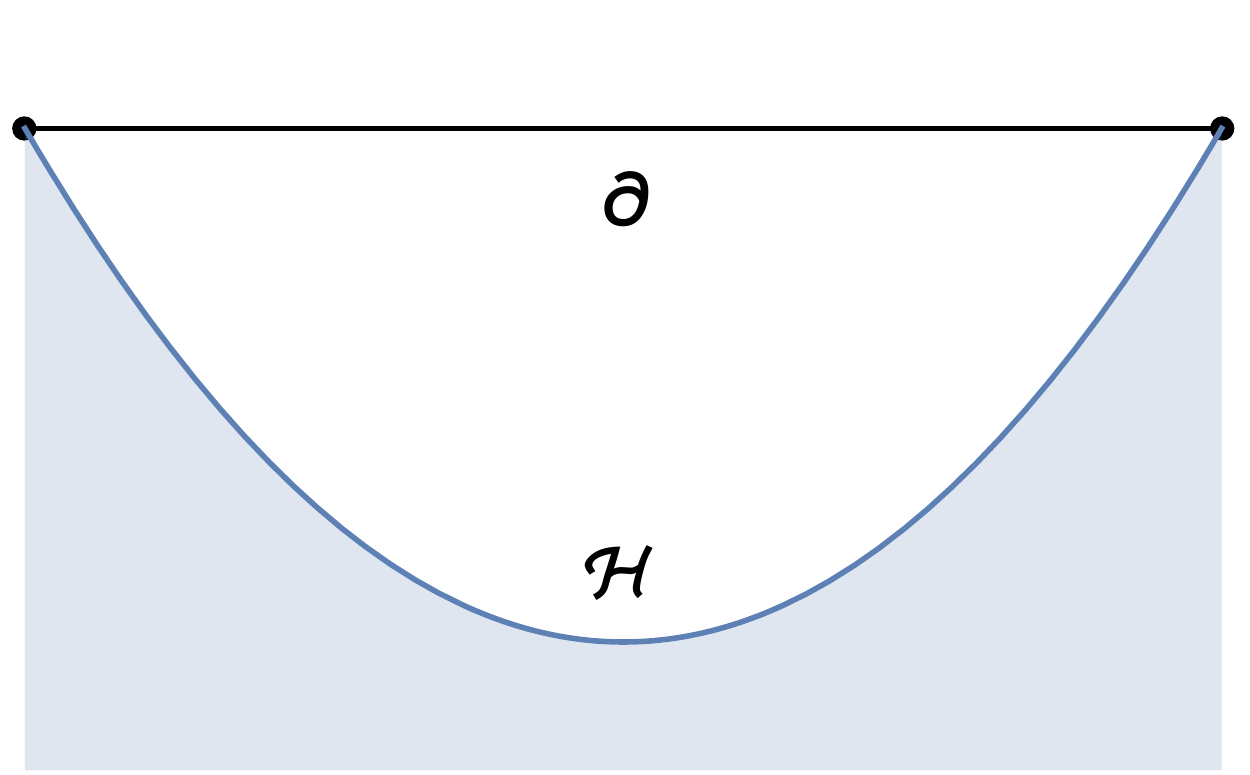}
  \includegraphics[width=7cm]{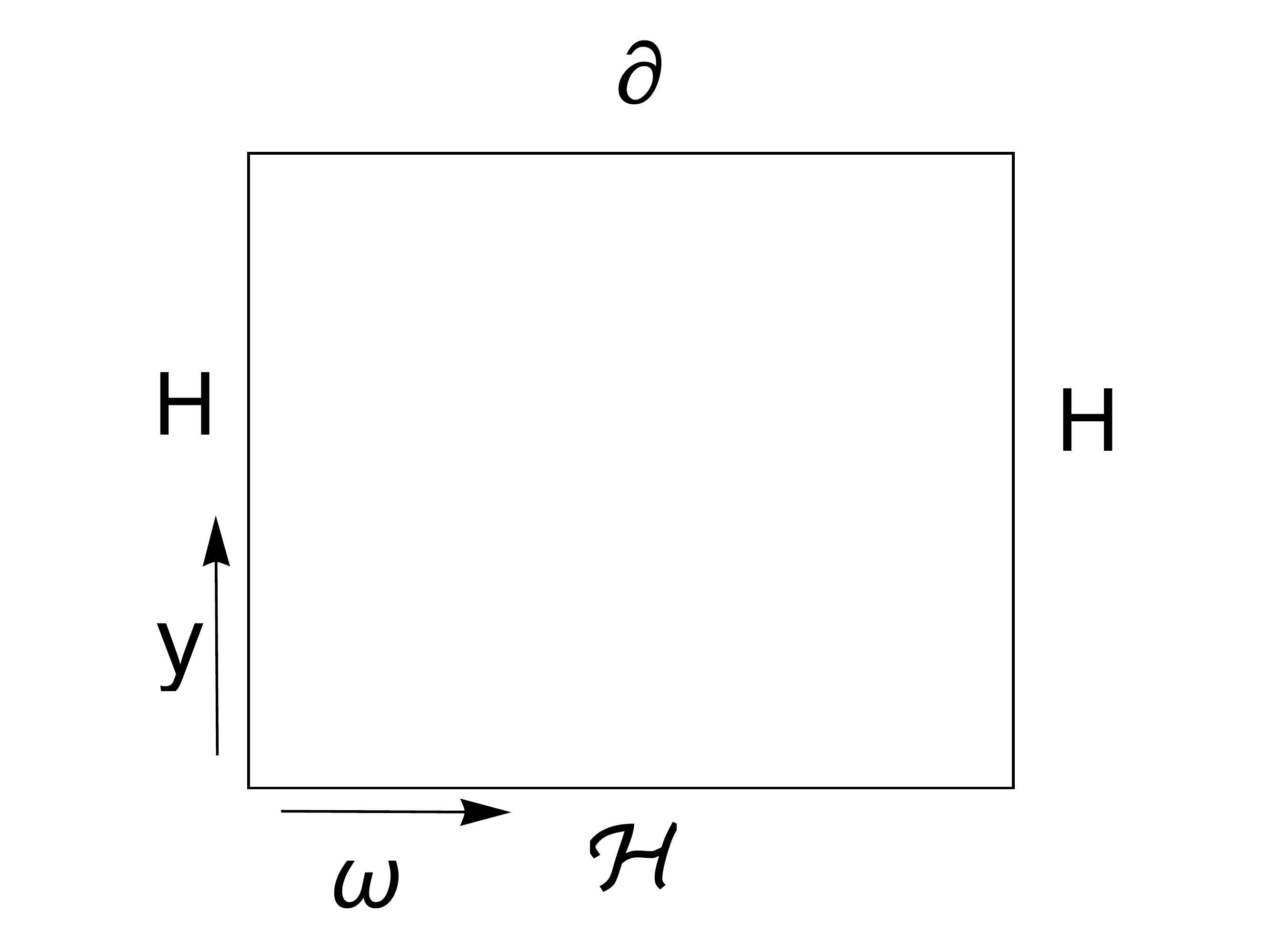}\\
  \caption{A sketch of black funnel solution. $\partial$ represents the AdS conformal boundary, $\mathcal{H}$ the bulk horizon, and $H$ the hyperbolic black hole. $\omega$ is the space coordinate along the bulk horizon and $y$ is the AdS radial coordinate.}\label{fig}
\end{figure}

In order to find a solution that describes the heat transport on the AdS boundary, one naturally consider that there are two heat reservoirs on the AdS boundary \cite{HMR}. The event horizon, which has a temperature from the thermodynamics of black hole, can play the role of the heat reservoir. A sketch of the black funnel solution is presented in Fig.\ref{fig}. The bulk horizon $\mathcal{H}$ of the black funnel solution meets the AdS conformal boundary $\partial$ at the black points which are the induced horizons on the AdS boundary as shown on the left panel of Fig.\ref{fig}.
In other words, there are two horizons (represented by two black points) on the AdS conformal boundary and the two horizons on the AdS boundary are connected through the horizon $\mathcal{H}$ in the bulk.

As shown on the left panel, the computational domain is not regular, which is not convenient for the numerical calculation. However, as shown on the right panel of Fig.\ref{fig}, the points where the bulk horizon $\mathcal{H}$ meet the AdS conformal boundary $\partial$ can be blown up into two different hyperbolic black holes at the left and right boundaries \cite{FMS}. In fact, this is the case for the region where the bulk horizon meets the boundary horizon. In this way, the computational domain was transformed into a rectangle. Therefore, the black funnel is constructed as a solution to the Einstein equations that has one bulk horizon $\mathcal{H}$, approaches two different hyperbolic black holes on either ends of this horizon, and has a boundary that is conformal to the background metric of the field theory. In addition, the left or right asymptotic hyperbolic black hole is fixed by the choice of the bulk horizon temperature and the boundary horizon temperature.
The two hyperbolic black holes with different temperatures can be regarded as the thermal reservoirs. Such a black funnel solution describes the heat flow between two reservoirs at different temperatures. In the following, we will show that the Hawking temperature and the chemical potential of the bulk horizon vary with the coordinate $\omega$, or vary along the bulk horizon. This implies that there are heat flow as well as particle flow driven by the difference in temperatures and chemical potentials along the bulk horizon.

 The metric of black funnel solution is given by \cite{FMS}
 \begin{eqnarray}\label{metric}
 ds^2=&&\frac{l_4^2}{(1-\omega^2)^2(1-y^2)^2}\left\{-M(y)G(\omega)^2(1-\omega^2)^2 y^2 A
 \left[l_4^{-1}dt+Q(\omega)\frac{\chi_2}{y}dy\right]^2\right.
 \nonumber\\&&
 \left.+\frac{4(1-\omega^2)^2Bdy^2}{M(y)}+
 y_0^2\left[\frac{4S_1}{(2-\omega^2)}\left(d\omega+l_4^{-1}\chi_1dt+\frac{Fdy}{y}\right)^2
 +S_2d\phi^2\right]\right\}\;,
 \end{eqnarray}
 where $A, B, F, S_1, S_2, \chi_1$ and $\chi_2$ are all functions of $y$ and $\omega$. In addition, other metric functions are defined as
 \begin{eqnarray}
 G(\omega)&=&1+\frac{\beta}{2}\omega^3(5-3\omega^2)\;, \nonumber\\
 M(y)&=&2-y^2-\frac{(1-y^2)^2(1-y_0^2)}{y_0^2}\;,\nonumber\\
 Q(\omega)&=&1-\frac{2}{M(0)G(\omega)}\;.
 \end{eqnarray}

 The parameter $l_4$ is the length scale of $AdS_4$ space. $y_0$ is a parameter
 that controls the validity of the fluid approximation. Here $y$ ranges over $[0,1]$ and $\omega$ ranges over $[-1,1]$, with $y=0$ being the bulk horizon and $y=1$ the conformal boundary.
 The black funnel solution is a solution that has a horizon, approaches two different hyperbolic black holes on either ends of this horizon, and has a boundary that is conformal to the background metric of the field theory. With the proper boundary conditions, the induced boundary metric has two horizons (two black holes). The above set of considerations is graphed in Fig. \ref{fig}.
 The parameter $\beta$ controls the temperature difference between the two boundary black holes. By AdS/CFT, this black funnel solution is used to describe the non-equilibrium field theory on the boundary.

 The above metric of the black funnel is considered to solve the vacuum Einstein equations with negative cosmological constant in the form of
 \begin{eqnarray}
 R_{\mu\nu}+\frac{3}{l_4^2}g_{\mu\nu}=0\;.
 \end{eqnarray}
 However, boundary conditions should be imposed when solving the Einstein equations. For our purpose,
 we only give the boundary conditions that imposed on the bulk horizon since the Hawking radiation is relevant to the near horizon properties of black hole spacetime.
By demanding the regularity of the black funnel metric in ingoing Eddington-Finkelstein coordinates (which cover the future horizon), the boundary conditions at the bulk horizon $y=0$ are taken as
 \begin{eqnarray}
 F(0,\omega)=\chi_1(0,\omega)\;, B(0,\omega)=\frac{1}{4}M(0)^2G(\omega)^2 A(0,\omega)
 \left[1-Q(\omega)\chi_2(0,\omega)\right]^2\;,\nonumber\\
 \partial_y A(0,\omega)=\partial_y S_1(0,\omega)=
 \partial_y S_2(0,\omega)=\partial_y \chi_1(0,\omega)=
 \partial_y \chi_2(0,\omega)=0\;.
 \end{eqnarray}
 With the metric ansatz and boundary conditions of the black funnel, the numerical solution can be found by using the Einstein-De Turk method \cite{DSW}. It should be noted that these boundary conditions will be
 supplemented to simplify the equation of motion of a scalar field near the horizon.

 Let $g$ denote the determinant of the metric of the black funnel. It can be calculated that
 \begin{eqnarray}
 \sqrt{-g}=\frac{4l_4^3 y_0^2 y}{(1-\omega^2)^2(1-y^2)^4}\sqrt{\frac{ABS_1S_2G(\omega)}{(2-\omega^2)}}\;.
 \end{eqnarray}
 It is obvious that the determinant of the metric is degenerate at the bulk horizon $y=0$. In the
 following, we set $l_4=1$ without the loss of generality.

 \subsection{Local Hawking temperature and local chemical potential of the nonequilibrium black hole-black funnel}

 To derive the spectrum of Hawking radiation, for simplicity, we consider the dynamics of a massive scalar field which is govern by the Klein-Gorden equation
 \begin{eqnarray}
 \nabla_{\mu}\nabla^{\mu}\psi=\frac{1}{\sqrt{-g}}
 \partial_{\mu}\left(\sqrt{-g}g^{\mu\nu}\partial_{\nu}\psi\right)=m^2\psi\;,
 \end{eqnarray}
 where $m$ denotes the mass of scalar field. The black funnel metric $g_{\mu\nu}$ are independent of the coordinates $t$ and $\phi$, which means that $\partial_t$ and $\partial_\phi$ are killing vector respectively. According to these symmetries, the Klein-Gorden equation can be rewritten in the following form
 \begin{eqnarray}
 &&g^{tt}\partial_t^2\psi
 +2g^{ty}\partial_t\partial_y\psi
 +2g^{t\omega}\partial_t\partial_\omega\psi
 +\frac{1}{\sqrt{-g}}\partial_y \left[\sqrt{-g}g^{ty}\right]\partial_t\psi
 +\frac{1}{\sqrt{-g}}\partial_y \left[\sqrt{-g}g^{yy}\partial_y\psi\right]
  \nonumber\\&&
 +\frac{1}{\sqrt{-g}}\partial_y \left[\sqrt{-g}g^{y\omega}\right]\partial_\omega\psi
 +2g^{y\omega}\partial_y\partial_\omega\psi
 +\frac{1}{\sqrt{-g}}\partial_\omega \left[\sqrt{-g}g^{t\omega}\right]\partial_t\psi
 +\frac{1}{\sqrt{-g}}\partial_\omega \left[\sqrt{-g}g^{y\omega}\right]\partial_y\psi
 \nonumber\\&&
 +\frac{1}{\sqrt{-g}}\partial_\omega \left[\sqrt{-g}g^{\omega\omega}\right]\partial_\omega\psi
 +g^{\omega\omega}\partial_\omega\psi
 +g^{\phi\phi}\partial_\phi^2\psi-m^2\psi=0\;.
 \end{eqnarray}

 We can introduce the generalized tortoise coordinate in the form of \cite{ZD}
 \begin{eqnarray}
 y*=\ln |y|=\left\{
              \begin{array}{ll}
                \ln y, & y>0 \\
                \ln(-y), & y<0
              \end{array}
            \right.
 \end{eqnarray}
 where $y>0$ denotes the spacetime outside the bulk horizon and $y<0$ denotes the spacetime inside the
 bulk horizon. By performing the generalized tortoise coordinate transformation, multiplying the Klein-Gorden equation by the factor $y^2$, and taking the limit
 $y\rightarrow 0$ and $\omega\rightarrow \omega_0$, where $\omega_0$ is an arbitrary spatial location
 in the $\omega$ direction, the resulting equation in generalized
 tortoise coordinate can be written as
 \begin{eqnarray}
 C_{tt} \partial_t^2\psi+C_t \partial_t \psi+C_{ty}\partial_r\partial_{y*}\psi
 +C_{t\omega}\partial_r\partial_{\omega}\psi+C_{yy}\partial_{y*}\partial_{y*}\psi
 +C_{y}\partial_{y*}\psi+C_{y\omega}\partial_{y*}\partial_{\omega}\psi=0\;,
 \end{eqnarray}
 where the coefficients are given by
 \begin{eqnarray}
 C_{tt}&=&\frac{1}{4M(0)}\left[\frac{M(0)^2Q(\omega_0)^2\chi_2(0,\omega_0)^2}{B(0,\omega_0)}
 -\frac{4}{A(0,\omega_0)G(\omega_0)^2}\right]\;,\nonumber\\
 C_{t}&=&\frac{\chi_1(0,\omega_0)\left(\partial_y \ln S_1(0,\omega_0)+\partial_y \ln S_2(0,\omega_0)+\frac{2\omega_0(9-5\omega_0^2)}{2-3\omega_0^2+\omega_0^4}\right)
 +2\partial_y\chi_1(0,\omega_0)}{2A(0,\omega_0)G(\omega_0)^2M(0)\left(1-Q(\omega_0)\chi_2(0,\omega_0)
 \right)}\nonumber\\
 C_{ty}&=&-\frac{M(0)Q(\omega_0)\chi_2(0,\omega_0)}{2B(0,\omega_0)}\;,\nonumber\\
 C_{t\omega}&=&\frac{M(0)\chi_1(0,\omega_0)\left(1-Q(\omega_0)\chi_2(0,\omega_0)\right)}{2B(0,\omega_0)}
 \;,\nonumber\\
 C_{yy}&=&\frac{M(0)}{4B(0,\omega_0)}\;,\nonumber\\
 C_{y}&=&-C_{t}\;,\nonumber\\
 C_{y\omega}&=&-C_{t\omega}\;.
 \end{eqnarray}

 The limit $y\rightarrow 0$ and $\omega\rightarrow \omega_0$ means that we only concentrate on the
 local behavior of the scalar field near the horizon and at the location of $\omega=\omega_0$. In this
 limit, the resulting radiation spectrum and Hawking temperature will depend on the spatial location
 $\omega_0$, which is very different from the results of the equilibrium black hole with killing horizon. This is an example of local temperature. We will show that Damour-Ruffini method is robust in deriving the Hawking radiation from the non-killing horizon of the black funnel.

 By performing the separation of variables, we take the ansatz of the scalar field as
 \begin{eqnarray}
 \psi=e^{-iEt}R(y*)e^{i k_\omega \omega + i k_\phi \phi}\;.
 \end{eqnarray}
 Substituting this expression into the reduced scalar equation, we can obtain the following
 radial equation after some tedious algebra
 \begin{eqnarray}
 L_2 \partial_{y*}^2 R+L_1 \partial_{y*} R + L_0 R=0\;,
 \end{eqnarray}
 with
 \begin{eqnarray}
 L_2&=&\frac{M(0)}{4B(0,\omega_0)}\;,\nonumber\\
 L_1&=&\frac{i E M(0)Q(\omega_0)\chi_2(0,\omega_0)}{2B(0,\omega_0)}
 -\frac{ik_\omega M(0)\chi_1(0,\omega_0)\left(1-Q(\omega_0)\chi_2(0,\omega_0)\right)}{2B(0,\omega_0)}
 -C_{t}\;,\nonumber\\
 L_0&=&\frac{E^2 M(0)(1-2Q(\omega_0)\chi_2(0,\omega_0))}{4B(0,\omega_0)}
 +\frac{E k_\omega M(0)\chi_1(0,\omega_0)\left(1-Q(\omega_0)\chi_2(0,\omega_0)\right)}{2B(0,\omega_0)}
 -i E C_{t}\;.
 \end{eqnarray}

 It is obvious that this equation has two independent solutions, which are given explicitly as
 \begin{eqnarray}
 &&R_{in}=e^{-iE y^*}\;,\\
 &&R_{out}=e^{\left(i E (1-2Q(\omega_0)\chi_2(0,\omega_0))-2i k_\omega \chi_1(0,\omega_0)(1-Q(\omega_0)\chi_2(0,\omega_0))-4 B(0,\omega_0)C_{t}/M(0)\right)y* }\;.
 \end{eqnarray}
 Then, the waves corresponding, respectively, to in-going wave and out-going wave on the horizon are
 \begin{eqnarray}
 &&\psi_{in}=e^{-iE v}e^{i k_\omega \omega + i k_\phi \phi}\;,\\
 &&\psi_{out}=e^{-iE v}e^{2 i (E-k_\omega \chi_1(0,\omega_0))(1-Q(\omega_0)\chi_2(0,\omega_0))y*}
 e^{-4 B(0,\omega_0)C_{t}/M(0)y*}e^{i k_\omega \omega + i k_\phi \phi}\;,
 \end{eqnarray}
 with $v=t+y*$ being defined as the ingoing Eddington-Finkelstein coordinate.

 The ingoing wave is analytic while the outgoing wave is not at the bulk horizon.
 There is a logarithmic singularity for the out-going wave solution at the bulk horizon.
 Damour and Ruffini suggested that the outgoing wave can be analytically extended through
 the lower half complex $y$ plane into the inside of the horizon \cite{DR}. In our present case,
 the analytical continuation to extend the outgoing wave outside the bulk horizon to the outgoing wave inside the horizon can be realized by making the replacement of
 \begin{eqnarray}
 y\mapsto |y|e^{-i\pi}=(-y)e^{-i\pi}\;.
 \end{eqnarray}
 Then, the outgoing wave inside the bulk horizon by analytical continuation can be given by
 \begin{eqnarray}
 \psi_{out}(y>0) \mapsto \psi_{out}(y<0)&=&e^{-iE v}e^{2 i (E-
 k_\omega \chi_1(0,\omega_0))(1-Q(\omega_0)\chi_2(0,\omega_0))y*}
 e^{-4 B((0,\omega_0))C_{t}/M(0)y*}e^{i k_\omega \omega + i k_\phi \phi}\nonumber\\&&
 e^{2 \pi (E-k_\omega \chi_1(0,\omega_0))(1-Q(\omega_0)\chi_2(0,\omega_0))}
 e^{4 i\pi B(0,\omega_0)C_{t}/M(0)}\;,
 \end{eqnarray}
 The relative probability of the scattered outgoing wave at the bulk horizon is given by
 \begin{eqnarray}
 P=\left|\frac{\psi_{out}(y>0)}{\psi_{out}(y<0)}\right|^2=e^{-4 \pi (E-k_\omega \chi_1(0,\omega_0))(1-Q(\omega_0)\chi_2(0,\omega_0))}\;.
 \end{eqnarray}
 According to the derivation of Sannan \cite{Sannan}, we obtain the distribution function of the outgoing
 energy flux (i.e. the spectrum of Hawking radiation from the bulk horizon)
 \begin{eqnarray}\label{radiation_Spectrum}
 N_{E}=\frac{1}{e^{(E-\mu)/T}-1}\;,
 \end{eqnarray}
 where the Hawking temperature of the black funnel is given by
 \begin{eqnarray}\label{Hawking_tep}
 T=\frac{1}{4\pi(1-Q(\omega_0)\chi_2(0,\omega_0))}\;,
 \end{eqnarray}
 and the chemical potential is given by
 \begin{eqnarray}\label{Local_Chem_potential}
 \mu=k_\omega \chi_1(0,\omega_0)\;.
 \end{eqnarray}

 As seen, both the local temperature $T$ and chemical potential $\mu$ are not constants along the bulk horizon. This indicates black funnel with non-killing horizon can not be in thermodynamic equilibrium state from the nonuniform temperature and chemical potential. In fact, the nonuniform temperature and chemical potential will generate heat flow and particle flow, respectively. This will leads to nonequilibrium thermodynamic dissipation. The corresponding nonequilibrium thermodynamics and evaporation dynamics of black funnel will be discussed in the next section.

 \subsection{Asymptotic behavior of the local Hawking temperature along the bulk horizon}

 Now, we consider the asymptotic behavior of the local Hawking temperature in the $\omega$ direction and compare with the temperatures of the asymptotic black holes. By taking the limit of $\omega_0\rightarrow \pm1$, the Hawking temperature takes the forms of
 \begin{eqnarray}
 T|_{\omega_0\rightarrow \pm 1}=\frac{M(0)G(\pm 1)}{8\pi}\;.
 \end{eqnarray}
 In the following, we can check this result by deriving the temperatures from the metric of the black funnel near the left and right boundaries. The left and right boundaries lie at $\omega=\pm 1$. There, the boundary conditions are imposed as
 \begin{eqnarray}
 &&A(\pm 1,y)=B(\pm 1,y)=S_1(\pm 1,y)=S_2(\pm 1,y)=\chi_2(\pm 1,y)=1,\nonumber\\
 &&F(\pm 1,y)=\chi_1(\pm 1,y)=0\;.
 \end{eqnarray}
 The metric in Eq.(\ref{metric}) is reduced to
 \begin{eqnarray}
 ds^2|_{\omega\rightarrow \pm 1}&=&\frac{1}{(1-y^2)^2}\left\{
 -M(y)G(\pm 1)^2 y^2\left[dt+Q(\pm 1)\frac{dy}{y}\right]^2+\frac{4dy^2}{M(y)}
 \right.\nonumber\\
 &&\left.+\frac{y_0^2}{(1\mp \omega)^2}\left(d\omega^2+\frac{d\phi^2}{4}\right)
 \right\}\;.
 \end{eqnarray}
 By coordinate transformation, it can be shown that this metric is asymptotic to the hyperbolic black hole. The temperature of this metric can be found by using the conventional method for the black hole with killing horizon (for example, using the definition of surface gravity and the relation between temperature and surface gravity $T=\frac{\kappa}{2\pi}$)
 \begin{eqnarray}
 T_{\pm}=\frac{M(0)G(\pm 1)}{8\pi}\;,
 \end{eqnarray}
 which is exactly the limit of the Hawking temperature of the black funnel. Therefore we show that the limits of the local Hawking temperature of bulk horizon are consistent with the Hawking temperatures of black holes on the left and right boundaries.

\subsection{Implications of the nonequilibrium thermodynamics for the boundary CFT}

 At last, let us discuss some implications of the local temperature of black funnel for the nonequilibrium thermodynamics of dual conformal field theory. The nonequilibrium black funnel solution was proposed to describe the far from equilibrium transport of heat on the boundary field theory \cite{FMS}. If the local Hawking temperature of the bulk horizon is identified with the local temperature of boundary field theory, the entropy production of the boundary field theory can be computed from the point view of AdS/CFT correspondence. In \cite{FMS}, due to the absence of the definition of the local temperature of the boundary field theory, the entropy production of the boundary conformal field was not calculated. However, the entropy current and its divergence can be computed by using the hydrodynamic approximation. With the local Hawking temperature given by Eq.(\ref{Hawking_tep}), it is possible to calculate the entropy current $\vec{J}$ and the local entropy production $\Theta$ for the conformal fluid by identifying the local temperature of the conformal fluid as the local temperature of the black funnel horizon. In non-equilibrium thermodynamics, the entropy current and entropy production rate are given by
 \begin{eqnarray}
 \vec{J}_S&=&\frac{\vec{J}_q}{T}\;,\nonumber\\
 \Theta&=&\vec{J}_q\cdot \nabla\left(\frac{1}{T}\right)\;.
 \end{eqnarray}
 where the energy current $\vec{J}_q$ can be determined by the metric of AdS funnel solution by employing the holographic renormalization technique. The obtained numerical results can be compared with the analytical results under the hydrodynamic approximation. This question depending on the numerics deserves further study in the future.

 \section{Nonequilibrium thermodynamics of black hole: black funnel}

 The conventional black holes with the killing horizons can be described by the equilibrium thermodynamics for its macroscopic emergent state. As discussed, the black funnel solution, which has a temperature distribution on its bulk horizon, should be treated as a non-equilibrium thermodynamic system. Therefore, the non-equilibrium thermodynamics is expected to describe the black funnel solution. It should be noted that the black funnel solution is time independent. This implies that black funnel is in a steady state. Due to the temperature variation, black funnel is in nonequilibrium steady state. Then the nonequilibrium thermodynamics of the black funnel should be dramatically different from the equilibrium thermodynamics of black hole solution with the killing horizon. In thermodynamics of the non-equilibrium steady state, the energy current, the entropy current, and the local entropy production are the characteristic quantities that describe the physical natures of the system. These quantities turn out to be closely related with the first and the second laws of non-equilibrium thermodynamics. The energy current, entropy current, as well as the local entropy production of the bulk horizon of the black funnel solution can also be quantified.

 \subsection{Nonequilibrium thermodynamics}

 To proceed with the nonequilibrium thermodynamics description of the black holes with non-killing horizon (black funnels), let us divide the black hole into smaller parts with each part still in equilibrium characterized by the local temperature and chemical potential although the overall system is in non-equilibrium. If the relaxation time of the parts are much faster than that of the whole system, the parts can be regarded as in local equilibrium.
 The local equilibrium can also be understand as follows. Between the microscopic and macroscopic emergent black hole, there is a mesoscopic scale with many microscopic degrees of freedom but not macroscopic degrees of freedom. So mesoscopic degrees of freedom is much larger than the microscopic degrees of freedom but much smaller than the macroscopic degrees of freedom. The local equilibrium is achieved in the mescopic scale. The whole macroscopic emergent scale of black hole is in non-equilibrium state with many local patches at their own local temperatures and chemical potentials on the horizon. The local entropy therefore refers to the mesoscopic degrees of freedom of the local patch.

 In this situation, the temperature, chemical potential, internal energy, entropy, and particle number can have the definitive meanings. We can assume that these local thermodynamic quantities still satisfy the thermodynamic first law. Dividing by volume, we can reach \cite{GM}
 \begin{eqnarray}
 Tds=du-\Sigma \mu_i d n_i\;,
 \end{eqnarray}
 where $s,u,n_i$ are the entropy density, local energy density, and particle density, and $T$ and $u$ are local temperature and chemical potential respectively. Here, we should explain the meaning of the volume. For Schwarzchild black hole, we have $dV=r dA$ where $dV$ is the small volume element while $r$ is the horizon radius and $A$ is the area of the horizon. For the black funnel, $dV=r dA$, everything has the same meaning except that $r$ now is varying and not a constant any more as in Schwarzchild black hole.

 In a fixed volume element, the change of the particle number density satisfies a local conservation equation of matter
 \begin{eqnarray}
 \frac{\partial n}{\partial t}=-\nabla\cdot J_n\;.
 \end{eqnarray}
 The physical meaning is clear: the change of the particle number density is equal to the net particle flow or flux density in or out of the volume. Similarly, the change of the internal energy density also satisfies the local conservation equation of energy
 \begin{eqnarray}
 \frac{\partial u}{\partial t}=-\nabla\cdot J_{u}\;.
 \end{eqnarray}
 The physical meaning is also clear: the change of the internal energy density is equal to the net energy flux density in or out of the volume.

 From the first law of the thermodynamics of the local volume, $Tds=du-\Sigma \mu_i d n_i$,
 we know that when particle number density is increased as $dn$, the internal energy change is $\mu dn$. Therefore, when the particle flow is present, the internal energy density flux can be quantified as
 \begin{eqnarray}
 J_{u}=J_{q}+\mu J_n\;.
 \end{eqnarray}
 Thus, the internal energy flux density is the sum of the heat flux density and particle flux density.
 By substituting the above formula into the local conservation law of the internal energy density, we reach \begin{eqnarray}
 \frac{\partial u}{\partial t}=-\nabla\cdot J_q-\nabla\cdot(\mu J_n)\;.
 \end{eqnarray}

 On the other hand, from the first law of equilibrium thermodynamics of the local volume $Tds=du-\mu dn$, we obtain
 \begin{eqnarray}
 \frac{\partial s}{\partial t}=\frac{1}{T}\frac{\partial u}{\partial t}-\mu \frac{\partial n}{\partial t}\;.
 \end{eqnarray}

 By substituting the local conservation law of the particle number density into the above expression, we get the change of the system entropy \cite{GM}
 \begin{eqnarray}
 \frac{\partial s}{\partial t}&=&-\frac{1}{T} \nabla\cdot J_q-\frac{1}{T}\nabla\cdot(\mu J_n)
 +\frac{\mu}{T}\nabla\cdot J_n\nonumber\\
 &=&-\nabla\cdot\left(\frac{J_q}{T}\right)+J_q\cdot \nabla\frac{1}{T}-\frac{J_n}{T}\cdot\nabla\mu\;.
 \end{eqnarray}
 The physical meaning is also clear: $\frac{\partial s}{\partial t}$ represents the change of the entropy of the system; $-\nabla\cdot\left(\frac{J_q}{T}\right)$ represents the change of the local entropy density due to the heat flow; $J_q\cdot\nabla\frac{1}{T}$ represents the local entropy density production rate due to the temperature gradient which leads to the heat transport; $-\frac{J_n}{T}\cdot\nabla\mu$ represents the local entropy production rate due to the chemical potential gradient which leads to the particle transport.
 It is well knownfrom the classical general relativity that only the photons can propagate along the event horizon. Here, the particles that transport along the horizon are the bosons with both temperature and chemical potentials. They are not like photons completely but rather act like Bose-Einstein gas (temperature and chemical potential). These particles will be moving on the horizon surface under the temperature and chemical potential gradient creating energy or heat flows driven by temperature gradient as well as the particle flows driven by the chemical potential gradient.

 We can define the entropy flux density and the total entropy production rate as
 \begin{eqnarray}
 J_s=\frac{J_q}{T}\;\;, epr=J_q\cdot\nabla\frac{1}{T}-\frac{J_n}{T}\cdot\nabla\mu\;.
 \end{eqnarray}
 We can also define the driving force generated by the nonuniform temperature and chemical potential as
 \begin{eqnarray}
 X_q=\nabla\frac{1}{T}\;\;,X_n=-\frac{1}{T}\nabla\mu\;,
 \end{eqnarray}
 where $X_q$ is called the driving force of the heat flow while $X_n$ ia called the driving force of the particle flow. Thus the local entropy density production rate can be expressed as the sum of the product of the driving force and the flux of the two kinds, one for the heat and the other for the particle.

 The equation for the entropy change in time as: $\frac{\partial s}{\partial t}=-\nabla\cdot\left(\frac{J_q}{T}\right)+J_q\cdot \nabla\frac{1}{T}-\frac{J_n}{T}\cdot\nabla\mu=-\nabla\cdot \left(\frac{J_q}{T}\right)+epr$ has a clear physical meaning: $\frac{\partial s}{\partial t}$ represents the change of the entropy of the system;
 $-\nabla\cdot\left(\frac{J_q}{T}\right)$ represents the change of the local entropy density due to the heat flow or from the environment; $epr=J_q\cdot \nabla\frac{1}{T}-\frac{J_n}{T}\cdot\nabla\mu$ represents the total local entropy production generated from the energy and the particle flows as well as the temperature and the chemical potential difference (system plus environment). In fact, this formulates the first law of the nonequilibrium thermodynamics. It can be shown that $epr\geq 0$. That is the total local entropy density change is always greater or equal to zero. This formulates the second law of the nonequilibrium thermodynamics.

 For a system, although the total local entropy density of the system and the environment together always prefer to increase leading to disorder of the whole system and environment, the local entropy density of the system can be decreased in the process.
 This is because that the heat or entropy flow to the system from the environment can be negative or the net positive heat or entropy flow from the system to the environment. The possibility of the decrease of the system entropy can create order. In fact, for living system, the entropy of the system is relatively low giving rise to order while the total entropy change still increases leading to disorder.

 \subsection{Nonequilibrium black hole thermodynamics}

 As discussed, if the black hole can be described by the normal thermodynamics for its macroscopic emergent state, then from the studies of the time dependent evaporation process and dynamical Vaidya black hole accretion process, it is natural to suggest that the nonequilibrium thermodynamics can also be used to describe the nonequilibrium process of the black hole.

 Notice that the nonequilibrium open system is very different from the equilibrium closed system where the entropy of the system always increases. For black hole with the non-killing horizon, such as black funnel in our study, we can write down the first law of its global nonequilibrium thermodynamics as
 \begin{eqnarray}\label{first_law}
 \frac{\partial S}{\partial t}=-\int \frac{J_q}{T}\cdot dA+\int J_q\cdot\nabla\frac{1}{T} dV
 -\int\frac{J_n}{T}\cdot\nabla\mu dV\;,
 \end{eqnarray}
 where $S$ is the total entropy of the black funnel, $dV$ is the volume element, and the integral is over the whole volume.

 $J_q$ is the heat energy flow density for the masslesss particles under chemical potential, and $J_n$ is the particle flux under the chemical potential gradient. 
 Here, we will present the formulas of particle flux of quantum gases driven by chemical potential gradient. The null nature of the event horizon of black holes restricts that only the massless particles can propagate along the horizon. In the present paper, we are particularly interested in the energy and particle fluxes along the horizon of black hole. Therefore, we only consider the case of massless particles. If including the fermionic field for black funnel, we expect Fermi-Dirac statistics under chemical potential for the Hawking radiation. In the following, we will consider the massless Bosons as well as the massless Fermions.   

For the massless particles, the energy can be expressed as $\epsilon=\hbar\omega$, where $\omega$ is the frequency. The number of states available to a single particle with frequency between $\omega$ and $\omega+d\omega$ is
\begin{eqnarray}
g(\omega)d\omega=\frac{V\omega^2d\omega}{\pi^2 c^3}\;.
\end{eqnarray}
According to Bose-Einstein/Fermi-Dirac distribution
\begin{eqnarray}
a_l=\frac{g_l}{e^{(\epsilon_l-\mu)/kT}\pm1}\;,
\end{eqnarray}
where $\pm$ is for fermion/Boson and $g_l$ is the degeneracy of the energy level $\epsilon_l$, we can easily compute the number density of the massless particles in the gas
\begin{eqnarray}
n(\mu)=\frac{N}{V}&=&\frac{1}{\pi^2 c^3}\int_{0}^{+\infty}\frac{\omega^2d\omega}{e^{(\hbar\omega-\mu)/kT}\pm1}\nonumber\\
&=&\frac{1}{\pi^2}\left(\frac{kT}{\hbar c}\right)^3\int_{0}^{+\infty}\frac{x^2 dx}{e^{-\mu/kT}e^x\pm 1}\;,
\end{eqnarray}
and the energy density for the massless particles
\begin{eqnarray}
u(\mu)=\frac{E}{V}&=&\frac{1}{\pi^2 c^3}\int_{0}^{+\infty}\frac{\omega^3d\omega}{e^{(\hbar\omega-\mu)/kT}\pm1}\nonumber\\
&=&\frac{1}{\pi^2 c^3}\left(\frac{kT}{\hbar}\right)^4\int_{0}^{+\infty}\frac{x^3 dx}{e^{-\mu/kT e^x}\pm1}
\;,
\end{eqnarray}

\begin{figure}
  \centering
  \includegraphics[width=7cm]{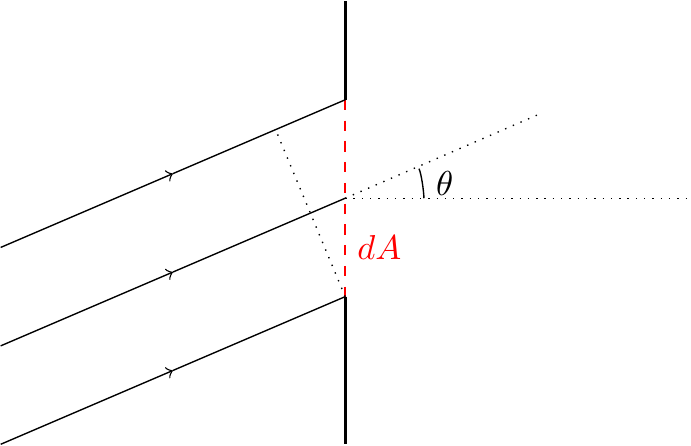}
  \caption{Energy and particle flux of Bose particles.  }
  \label{Energy_Flux}
\end{figure}

On the other hand, in radiation theory, there are formulas for energy flux and particle flux.  As shown in Fig.\ref{Energy_Flux}, the energy flow per unit time in certain direction through area $dA$ is equal to $c u\frac{d\Omega}{4\pi} \cos\theta dA$, where $d\Omega$ represents the solid angle and $\theta$ is the angle between the propagation and area $dA$. Integrating over all propagating directions, one can obtain the total energy flow as 
\begin{eqnarray}
 J_q dA=\frac{cu}{4\pi}\int \cos\theta d\Omega=\frac{1}{4} cu dA\;,
 \end{eqnarray}
Therefore, the well known formula for the energy flux of the radiation passing through a unit area in a unit time \cite{Huang} is given by 
\begin{eqnarray}
 J_q =\frac{1}{4} cu \;,
 \end{eqnarray}
 where $c$ is the speed of light and $u$ is the energy density.
 
 Correspondingly, the particles flow per unit time in certain direction through $dA$ is equal to $cn \frac{d\Omega}{4\pi} \cos\theta dA$, where $n$ is the particle number density. Integrating over all the directions, we have the particle flux defined as the particle number passing through a unit area in a unit time as 
  \begin{eqnarray}
 J_n=\frac{1}{4} cn\;,
 \end{eqnarray}
 where $n$ is the number density.
 If we treat the neighboring two volumes with different chemical potential $\mu$ and $\mu+\delta\mu$ are in local equilibrium states respectively, one can derive the particle flux between the two neighboring volumes as
\begin{eqnarray}
J_n=\frac{1}{4}c \left(n(\mu+\delta\mu)-n(\mu)\right)\;.
\end{eqnarray}
In principle, with the number density of the particles in hand, one can calculate the particle flux driven by the chemical potential gradient by using this formula.

 In Eq.(\ref{first_law}), $\nabla$ represents the gradient with respect to the location coordinates ($\omega$ here) on the black hole (black funnel) horizon. $dA$ represents the area element on the horizon. $\int dA$ represents the integral over the area of the horizon while $dV$ represents the integral over the black hole volume. $\nabla\frac{1}{T}$ represents the heat flow driving force in analogy to the voltage in the electric circuit where the heat flow density is in analogy to the electric current flow. Then $\int J_q\cdot\nabla\frac{1}{T} dV$ is the entropy production rate which represents the thermodynamic cost or dissipation due to the heat flow as the product of the heat flow and the heat flow driving force in analogy to the electric power generated by the product of the electric current and voltage. Since the temperature of the black funnel is given in Eq.(\ref{Hawking_tep}) as
 \begin{eqnarray}
 T(\omega)=\frac{1}{4\pi(1-Q(\omega)\chi_2(0,\omega))}\;,
 \end{eqnarray}
 depending only on one local coordinate, the heat driving force takes the gradient to $\omega$.
 In this case of black funnel, the local temperature only varies on the horizon surface. Thus the entropy production of $\int J_q\cdot\nabla\frac{1}{T} dV$ becomes an effective surface integral of
 $\int J_q\cdot\nabla\frac{1}{T} RdA$ where $R$ is the radius, in general not a constant on the black funnel horizon.

 Similarly, $J_n$ represents the particle flux while $\nabla\mu$ is the gradient of the chemical potential giving rise to the driving force for the particle flow. The product of the particle flow and the associated driving force $-\int\frac{J_n}{T}\cdot\nabla\mu dV$ represents the entropy production rate giving rise to the thermodynamic cost or the dissipation due to the particle flow. Similarly, the particle driving force $\nabla\mu$ also depends only on the local coordinate $\omega$ on the horizon as in Eq.(\ref{Local_Chem_potential}). Again, in this case of black funnel, the local chemical potential only varies on the horizon surface. Therefore, the entropy production $-\int\frac{J_n}{T}\cdot\nabla\mu dV$ becomes an effective surface integral.

 The first law of the nonequilibrium thermodynamics of the black hole can thus be formulated explicitly as the total entropy production being the sum of the entropy production of the system and the environment
 \begin{eqnarray}
 EPR=\frac{dS_{tot}}{dt}=\frac{dS}{dt}+\frac{dS_{env}}{dt}\;,
 \end{eqnarray}
  where $EPR=\frac{dS_{tot}}{dt}=\int J_q\cdot\nabla\frac{1}{T} dV
 -\int\frac{J_n}{T}\cdot\nabla\mu dV$, $\frac{dS}{dt}$ is the entropy change of the system, and
 $\frac{dS_{env}}{dt}=\int \frac{J_q}{T}\cdot dA$ is the entropy change of the environment.

 Since the total entropy production rate
 \begin{eqnarray}
 EPR=\frac{dS_{tot}}{dt}\geq 0\;,
 \end{eqnarray}
 this formulates the second law of the nonequilibrium thermodynamics of the black funnel. Note that at the steady state, the entropy change of the system is equal to zero, i.e. $\frac{dS}{dt}=0$. Then at the steady state, one has
 \begin{eqnarray}
 EPR=\frac{dS_{tot}}{dt}=\frac{dS_{env}}{dt}\;.
 \end{eqnarray}
 This indicates that at steady state the change of the total entropy is from the environment. In other words, to maintain a steady state of the nonequilibrium black hole, there is a thermodynamic cost to pay which is caused by the heat dissipation against the environments. We can see that $EPR$ can be used to measure the degree of nonequilibriumness away from the equilibrium.

 \subsection{Linear nonequilibrium thermodynamics of black hole-black funnel}

 As known, many nonequilibirum processes are due to certain inhomogeneity of the system. Temperature
 gradient can induce to the heat flow or energy flow while the concentration or density gradient can lead to particle transport or diffusion. Empirically, the current generated by such inhomogeneities is often relatively small, that is not very far from equilibrium:
 \begin{eqnarray}
 J=LX
 \end{eqnarray}
 where $J$ represents the transport current flow (mass, charge, momentum, heat or energy) per unit time per unit cross section area while $X$ represents the driving force (density gradient, voltage gradient, velocity gradient, temperature gradient).

 When several current flows and driving forces simultaneously are present, coupling behaviors emerge.
 The more general empirical linear nonequilibrium law can be formulated as
 \begin{eqnarray}
 J_k=\sum_{l} L_{kl}X_l
 \end{eqnarray}
  where $L_{kl}$ is the Onsager coefficient represents intensity of $k$th current caused by the $l$th driving force. The above relationship is often called Onsager relationship \cite{GM,Onsager}. In our case of the black funnel, both temperature gradient and chemical potential gradient on the horizon are present. As a result, both particle and thermal current emerge on the horizon. Therefore,
  assuming the linear relationshio between the current and the force not very far from equilibrium,
   the currents are approximately determined by the driving force as follows:
 \begin{eqnarray}
 J_q=-L_{21}\frac{1}{T}\nabla\mu+L_{22}\nabla\frac{1}{T}\nonumber\\
 J_n=-L_{11}\frac{1}{T}\nabla\mu+L_{12}\nabla\frac{1}{T}
 \end{eqnarray}
 According to Onsager reciprocal relationship \cite{Onsager}, $L_{12}=L_{21}$. As we see, for black funnel, the temperature gradient on the horizon not only can lead to thermal or energy current on the horizon, but also the particle transport on the horizon.

 In the same way, the chemical potential gradient on the horizon not only can induce the particle flow but also the thermal or energy flow on the horizon of the black funnel. This is due to the coupling   revealed in the generalized Onsager relations for linear nonequilibrium thermodynamics not too far from equilibrium. Therefore, $L_{11}$ represents the intensity of the thermal induced diffusion on the horizon. The temperature gradient on the horizon can lead to particle transport on the horizon through $L_{12}$. This phenomena is called Soret effect \cite{GM,HWKM}. $L_{21}$ represents the intensity of current induced by the thermal flow on the horizon. The chemical gradient can give rise to the density or concentration gradient and thus lead to the thermal current on the horizon through $L_{21}$. This phenomena is called Dufour effect \cite{GM,IH}. When the temperature gradient or the chemical gradient is small, the thermal induced diffusion or the diffusion induced thermal flow on the horizon will usually be small. However, when the temperature gradient or the chemical potential gradient becomes large, the thermal induced diffusion or the diffusion induced thermal current on the horizon can be significant. Since the total entropy production is given by
 \begin{eqnarray}
 epr=J_q\cdot \nabla \frac{1}{T}-\frac{J_n}{T}\nabla \mu\;.
 \end{eqnarray}
 We then have four contributions for the entropy production
 \begin{eqnarray}
 epr&=&\left(-L_{21}\frac{1}{T}\nabla\mu+L_{22}\nabla\frac{1}{T}\right)\cdot \nabla \frac{1}{T}
 \nonumber\\&&
-\left(-L_{11}\frac{1}{T}\nabla\mu+L_{12}\nabla\frac{1}{T}\right)\frac{\nabla \mu}{T}
\nonumber\\
&=&L_{11}\left(\frac{\nabla \mu}{T}\right)^2-L_{12}\nabla\left(\frac{1}{T}\right)^2\cdot \nabla \mu
+L_{22}\left(\nabla\frac{1}{T}\right)^2
 \end{eqnarray}
 $L_{11}\left(\frac{\nabla \mu}{T}\right)^2$ represents the contribution to the entropy production rate from the chemical potential gradient induced particle flow on the horizon.
 $L_{22}\left(\nabla\frac{1}{T}\right)^2$ represents the contribution to the entropy production rate from the temperature gradient induced thermal flow on the horizon.
 $-L_{12}\nabla\frac{1}{T}\cdot \frac{\nabla \mu}{T}$ representd the contribution to the entropy production rate from the thermal gradient induced particle current or diffusion on the horizon.
  $-L_{21}\nabla\mu\cdot\frac{1}{T}$ represents the contribution to the entropy production rate from the chemical potential gradient induced thermal current flow on the horizon. Note that the second law of the thermodynamics guarantees that the total entropy production rate $epr$ of the nonequilibrium black hole (the black funnel here) is larger than zero.

  When the temperature gradient and the chemical potential gradient are relatively small, the contributions to $epr$ from $L_{12}$ and $L_{21}$ parts are often less significant than the ones from $L_{11}$ and $L_{22}$ parts on the black funnel horizon. When the temperature and chemical potential gradients becomes high, contributions to $epr$ from $L_{12}$ and $L_{21}$ parts may not be ignored compared to the ones from $L_{11}$ and $L_{22}$ parts on the black funnel horizon.

 \subsection{Time arrow from the nonequilibrium black hole}

 The entropy production rate not only provides a measure of thermodynamic cost or dissipation for nonequilibrium black hole, but also gives a qualitative predictor of the degree of nonequilibriumness.

 One important implication of the non-zero entropy production rate is the time irreversibility or time arrow. From the fluctuation theorem, the ratio of the distribution of the state variable $C$ along the forward in time path $C(t)$ and backward in time path $\tilde{C}(t)$ is given by \cite{CrooksJSP,CrooksPRE}
 \begin{eqnarray}
 \frac{P[C(t)]}{P[\tilde{C}(t)]}=\exp [\Delta S_{tot}]\;,
 \end{eqnarray}
 where $\Delta S_{tot}$ is the entropy production. If there is a time reversal symmetry between backward and forward in time path, then the entropy production is zero. This is the conclusion for the equilibrium state with time reversal symmetry such as the conventional black hole with killing horizon. On the other hand, the nonzero entropy production implies the forward in time path has a different probability than than that in backward in time path. This means that the time reversal symmetry is broken by the nonzero entropy production. The nonzero entropy production thus gives rise to the time arrow. Therefore, we see that from the nonequilibrium boundary CFT, the corresponding bulk black hole is in the nonequilibrium steady state due to AdS/CFT correspondence. This nonequilibrium black hole in the bulk has a temperature and chemical potential variation or gradient on the horizon. It breaks the detailed balance and leads to the entropy production or dissipation cost. This dissipation provides the thermodynamic origin of the time asymmetry or time arrow.

 \section{Evaporation dynamics of the nonequilibrium black hole-black funnel}

 The radiation spectrum of the black funnel as presented in Eq.(\ref{radiation_Spectrum}) should be treated as the local property of the radiation field. The global spectrum from black funnel horizon that is in non-equilibrium state is expected to be different from the equilibrium Hakwing radiation spectrum. Consequently, the lifetime of the black funnel due to the non-equilibrium radiation will be very different from that of the equilibrium black hole.

 Let us consider the black hole evaporation dynamics. In the conventional equilibrium black hole evaporation approach, the energy loss which leads to the evaporation is purely from the Hawking radiation. In contrast, for the nonequilibrium black holes, such as black funnel, the evaporation process is not only determined by the Hawking radiation but also by the dissipation from the nonequilibriumness. By using the first law of the nonequilibrium thermodynamics of the black holes we just formulated, we can see
 \begin{eqnarray}\label{entropy_prod}
 \frac{dS}{dt}==-\int \frac{J_q}{T}\cdot dA+\int J_q\cdot\nabla\frac{1}{T} dV
 -\int\frac{J_n}{T}\cdot\nabla\mu dV\;,
 \end{eqnarray}
 where $S$ is the entropy of the black hole which is in turn given by the area of the black hole horizon.

 For the conventional equilibrium black hole, there is an uniform temperature and chemical potential on the horizon. Therefore the entropy productions involving the gradients of temperature and chemical potential vanishes. For the Schwarzschild black hole, $S\sim A\sim M^2$, $T\sim\frac{1}{M}$, and $J_q\sim T^4\sim \frac{1}{M^4}$. The $J_q\sim T^4$ assumes the flux in the cavity volume of the black body, it works for the photon energy flux. In principle the radiation can be directed to any directions not limited to the horizon surface. The reason why the entropy production is not effective outside the horizon is that the gradient of temperature or chemical potential is zero outside the horizon. So effectively the integration for the entropy production is over the horizon surface. Then the first law becomes
 \begin{eqnarray}\label{Sch_Rad}
 M\frac{dM}{dt}\sim -\frac{1}{M}\;.
 \end{eqnarray}
 This gives $M(t)\sim (1-\lambda t)^{\frac{1}{3}}$ which is exactly the same evaporation dynamics of the initial equilibrium black hole.

 We can see clearly that for an initially nonequilibrium black hole, in addition to the Hawing radiation leading to the energy or entropy loss, there is also a significant contribution from the nonequilibrium thermodynamic dissipation in term of the entropy production due to the non-uniform temperature and chemical potential. Since the entropy production is always positive, it is expected that the evaporation process of the nonequilibrium black hole tends to be slower than the initial equilibrium black hole with the Hawking radiation alone.

\begin{figure}
  \centering
  \includegraphics[width=7cm]{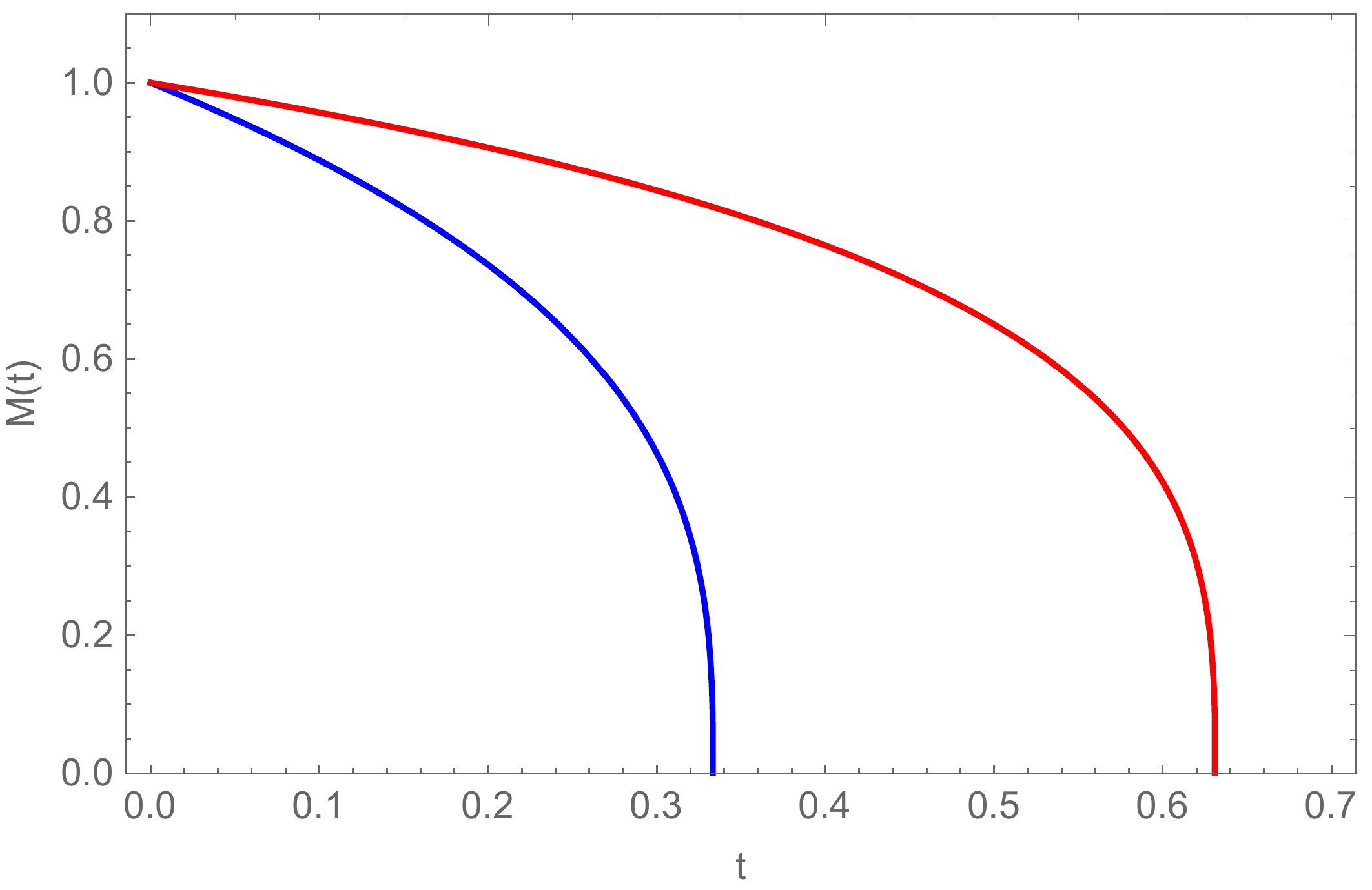}
  \caption{Illustration of Hawking evaporation process of the Schwarzschild black hole (blue line) and the black funnel (red line). The two curves are plotted based on the Eq. (50) and (52). The parameter $C$ is chosen as $0.6$. }\label{HR}
\end{figure}

 One can estimate this conclusion by using the first law of the nonequilibrium thermodynamics for the black funnel. We naively make a crude approximate assumption that the entropy and the mass of the black funnel still satisfy the relation of the entropy and the mass of the the Schwarzschild black hole, i.e. $S\sim M^2$. The second term $ \int J_q\cdot\nabla\frac{1}{T} dV$ on the right side of Eq.(\ref{entropy_prod}) is the entropy production due to temperature gradient $\nabla T$,  which is a positive number denoted as $C_1$. On the other hand, the entropy production has a contribution $C_2$ from the chemical gradient $-\int J_n\cdot \nabla\mu dV$ which is also a positive number. Let us define $C=C_1+C_2$. This term represents the total entropy production, which implies that $C>0$. If we consider the contribution of the heat flow from temperature gradient and the particle flow from chemical potential gradient to the entropy production, the first law gives the evolution equation for the black funnel as follows
 \begin{eqnarray}\label{BF_rad}
 M\frac{dM}{dt}\sim -\frac{1}{M}+C\;.
 \end{eqnarray}
 Based on the Eq. (\ref{Sch_Rad}) and (\ref{BF_rad}), we have plotted the mass of the Schwarzschild black hole as a function of time $t$ as well as the mass of the black funnel in Fig. \ref{HR}.
 It is seen that the evaporation time of the nonequilibrium black hole appears to be longer than that of the equilibrium black hole. In other words, the nonequilibrium effect is to slow down the evaporation process of the black funnel through the thermodynamic dissipation by the heat current flow from the
 temperature gradient and the particle current flow from the chemical potential gradient.

\section{Conclusion}

 In summary, we have derived the Hawking radiation from the bulk horizon of the non-equilibrium black funnel solution based on the Damour-Ruffini method. This provides the first derivation of Hawking temperature of the non-killing horizon as far as we know.
 Our results indicate that the spectrum, the temperature, and the chemical potential of the non-equilibrium black funnel do depend on one of the spatial coordinates. This is different from the conventional case of the constant temperature equilibrium black hole with killing horizon. Therefore, the black hole with non-killing horizon can be overall in nonequilibrium steady state while the Hawking temperature of this black funnel can be viewed as the local temperature and the corresponding Hawking radiation can be regarded as being in local equilibrium with the horizon of black funnel. Our results indicate that the black funnel horizon itself, which is a non-killing horizon, should be viewed as a non-equilibrium thermodynamics system.

 We further discuss the nonequilibrium thermodynamics of the black funnel, where the first law can be formulated as the entropy production rate being equal to sum of the changes of the entropies from the system and environments while the second law is given by the entropy production being larger than or equal to zero. We show that the nonequilibrium black hole leads to a time arrow due to the inhomogeneous distribution of the temperatures and the chemical potentials on the horizon giving rise to heat and particle transport dissipation. We also discuss how the nonequilibrium dissipation influences on the evaporation process of black funnel.

 \section*{Appendix A: Hawking radiation of black droplet}

In this appendix, we will compute the Hawking temperature of black droplet solution
by using the Damour-Ruffini method. Black droplet solution is another type of solution conjectured by Marolf et al \cite{HMR} to describe the gravity dual of heat transport between the boundary black hole. However, unlike the black funnel solution, the bulk horizon of black droplet is disconnected with the horizon of the boundary black hole, which suggests that the coupling between the boundary black hole
and the heat bath at infinity is suppressed. The black droplet solution describes the boundary CFT in equilibrium state and the temperature of droplet horizon should be constant. Now, we performed the computation of spectrum of Hawking radiation to verify this point.

 The metric of black droplet is given by \cite{SW}
 \begin{eqnarray}
 ds^2&=&\frac{1}{y h}\left\{-\frac{f_y f_\rho}{f_y+f_\rho-f_y f_\rho}T_c
 dt^2+\frac{A_c}{4y}dy^2\right.
 \nonumber\\&&\left.
 +\frac{B_c}{(1-x)^4}\left[dx+\frac{x(1-x)^3 F_c}{x^2+(1-x)^2y}dy\right]^2
 +\frac{x^2 S_c}{(1-x)^2}d\Omega_2^2
\right\}\;,
 \end{eqnarray}
 where $T_c, A_c, B_c, F_c$, and $S_c$ are functions of the Cartesian coordinates $x$ and $y$, and
 $d\Omega_2^2$ is the metric of two-dimensional unit sphere. $f_y$ and $f_\rho$ which are functions
 of $y$ and $\rho$ respectively, are given by
 \begin{eqnarray}
 f_y(y)&=&f(y)(1-\lambda y)^2\;,\nonumber\\
 f_\rho(\rho)&=&\frac{(1-\rho)^2}{(1+\rho)^6}\;,
 \end{eqnarray}
 with $f(y)$ being a smooth and positive definite function and the coordinate transformation
 \begin{eqnarray}
 y=\frac{R_0^2}{\rho^2}\left(1-\frac{x^2}{x^2+(1-x)^2y}\right)\;.
 \end{eqnarray}
 The function $h$ is also a smooth and positive definite function of $y$. $h(y)$ and $f(y)$ can be
 determined numerically by solving the Einstein equations. The conformal boundary locates at $y=0$,
 while the bulk horizon locates at $y=1/\lambda$. The boundary conditions at the bulk
 horizon $y=1/\lambda$ are given by
 \begin{eqnarray}
 T_c|_{y=1/\lambda}=A_c|_{y=1/\lambda}\;, F_c|_{y=1/\lambda}=0\;,\nonumber\\
 \partial_y T_c|{y=1/\lambda}=\partial_y A_c|{y=1/\lambda}=
 \partial_y B_c|{y=1/\lambda}=\partial_y S_c|{y=1/\lambda}=0\;.
 \end{eqnarray}

 The determinant of the matric is given by
 \begin{eqnarray}
 \sqrt{-g}=\frac{x^2 S_c \sin\theta}{2y^3(1-x)^4 h^2}
  \sqrt{\frac{A_cB_cT_cf_y f_\rho}{h(f_y+f_\rho-f_y f_\rho)}}\;.
 \end{eqnarray}
 We consider the Klein-Gorden equation for the scalar field in this background. It can be explicitly
 written as
 \begin{eqnarray}
 &&g^{tt}\partial_t^2\psi+\frac{1}{\sqrt{-g}}\partial_y\left[\sqrt{-g}g^{yy}\partial_y\psi\right]
 +\frac{1}{\sqrt{-g}}\partial_y\left[\sqrt{-g}g^{yx}\partial_x\psi\right]
 \nonumber\\
 &&+\frac{1}{\sqrt{-g}}\partial_x\left[\sqrt{-g}g^{xy}\partial_y\psi\right]
 +\frac{1}{\sqrt{-g}}\partial_x\left[\sqrt{-g}g^{xx}\partial_x\psi\right]
 +\frac{1}{\sqrt{-g}}\partial_\theta\left[\sqrt{-g}g^{\theta\theta}\partial_\theta\psi\right]
 \nonumber\\&&
 +g^{\phi\phi}\partial_\phi^2\psi-m^2\psi=0\;.
 \end{eqnarray}

 We introduce the generalized tortoise coordinate in the form of
 \begin{eqnarray}
 y*=\frac{1}{2\kappa}\ln |y-1/\lambda|=\left\{
              \begin{array}{ll}
                \frac{1}{2\kappa}\ln(1/\lambda- y), & 0<y<1/\lambda \\
                \frac{1}{2\kappa}\ln(y- 1/\lambda), & y>1/\lambda
              \end{array}
            \right.
 \end{eqnarray}
 where $\kappa$ is an adjustable parameter and will be determined in the following, $0<y<1/\lambda$ denotes the spacetime outside the bulk horizon and $y>1/\lambda$
 denotes the spacetime inside the bulk horizon.

 By multiplying the Klein-Gorden equation by $(1/\lambda-y)^2$, and taking the limit
 $y\rightarrow 1/\lambda$ and $x\rightarrow x_0$, where $x_0$ is an arbitrary spatial location
 in the $x$ direction, the resulting equation in generalized
 tortoise coordinate can be written as
 \begin{eqnarray}
C_{tt}\partial_t^2\psi+\frac{C_{yy}}{4\kappa^2}\partial_{y*}^2\psi=0\;,
 \end{eqnarray}
 with
\begin{eqnarray}
C_{tt}=\frac{h(1/\lambda)}{\lambda^3 f(1/\lambda)T_c(1/\lambda,x_0)}\;,\;C_{yy}=\frac{4h(1/\lambda)}{\lambda^2 A_c(1/\lambda,x_0)}\;.
\end{eqnarray}

 If choosing the parameter $\kappa=\frac{1}{2}\sqrt{\frac{C_{yy}}{C_{tt}}}=\sqrt{\lambda f(1/\lambda)}$, the above equation can be reduced to the standard wave equation as
 \begin{eqnarray}
 \partial_{y*}^2\psi-\omega^2 \psi=0\;.
 \end{eqnarray}
 According to the same method presented in Section II, one can obtain the Hawking temperature
 of black droplet solution as
 \begin{eqnarray}
 T=\frac{\sqrt{\lambda f(1/\lambda)}}{2\pi}\;.
\end{eqnarray}
 It is easy to see that the Hawking temperature of black droplet solution is constant.

 \end{document}